# An Evaluation of a Structured Spreadsheet Development Methodology

*Kamalasen Rajalingham, David Chadwick, Brian Knight*
*Information Integrity Research Centre*
*School of Computing & Mathematical Sciences*
*University of Greenwich*
*30 Park Row, Greenwich*
*London SE10 9LS, United Kingdom*

K.Rajalingham@gre.ac.uk

**ABSTRACT**

*This paper presents the results of an empirical evaluation of the quality of a structured methodology for the development of spreadsheet models, proposed in numerous previous papers by Rajalingham K, Knight B and Chadwick D et al. This paper also describes an improved version of their methodology, supported by appropriate examples. The principal objective of a structured and disciplined methodology for the construction of spreadsheet models is to reduce the occurrence of user-generated errors in the models. The evaluation of the effectiveness of the methodology has been carried out based on a number of real-life experiments. The results of these experiments demonstrate the methodology's potential for improved integrity control and enhanced comprehensibility of spreadsheet models.*

## 1 INTRODUCTION

This methodology represents a significant development or advance in the research into integrity control of spreadsheet models and the development of a methodology for spreadsheet model development. An outline of the methodology is presented by Knight B et al [Knight et al, 2000] and Chadwick D et al [Chadwick et al, 1999].

In this paper, an enhanced version of the methodology is presented. The methodology is essentially based on structured analysis of data, the outcome of which is represented as *Jackson-like structures*. It is shown that this analysis allows a straightforward modularisation, and that individual modules may be represented with indentation in the *block-structured* form of structured programs. The benefits of structured format are discussed, in terms of comprehensibility, ease of maintenance, and reduction in errors [Knight et al, 2000].

In order to assess and establish the quality of the methodology, four different experiments have been carried out. The results of these experiments have been analysed and they are presented in this paper.

## 2 A STRUCTURED METHODOLOGY FOR SPREADSHEET MODELLING

### 2.1 Introduction

Based on software engineering principles, mainly borrowed from Jackson [Jackson-1975], it has been found that spreadsheet models can be represented in a form identical to the data structure diagram developed by Jackson. Jackson [Jackson-1975] has shown how these *Structure Diagrams* can be mapped onto program code. In this paper, the proposed methodology demonstrates how these techniques can in fact be transferred to the production of spreadsheets, and this can give a more comprehensible format for spreadsheets, based on *indented* format. This is done by using a structured algorithm.

### 2.2 The Structured Algorithm Underpinning the Methodology

The algorithm consists of seven principal stages:

- Specification and Design of *Outputs*
- Conceptual Design of the *Workings* Section
- Logical Design of the *Workings* Section

- Construction of the Workings Section Structure
- Construction of the Input Section Structure
- Implementation of Functions and Relationships
- Completion of the Output Section

In this section, the methodology is applied in the construction of a spreadsheet model comprising a single module (as defined by the methodology). This is a simple model which does not require resolution of graph structures, which potentially result in the creation of separate modules, and recursive relationships. It is based on a *Trading and Profit and Loss Account* [Ward-1996]. The original model is shown in *Figure 1*.

```
                        T Howe Ltd
Trading and Profit and Loss Account for the year ended 31 December 19X4
Sales                                                     135,486
Less Cost of goods sold
    Opening stock                              40,360
    Add Purchases                              72,360
    Add Carriage inwards                        1,570
                                              114,290
Less Closing stock                             52,360     61,930
Gross profit                                              73,556
Less Expenses
    Salaries                                   18,310
    Rates and occupancy                         4,515
    Carriage outwards                           1,390
    Office expenses                             3,212
    Sundry expenses                             1,896
    Depreciation: Buildings                     5,000
                  Equipment                     9,000
    Directors' remuneration                     9,500     52,823
Net profit                                                20,733
Add Unappropriated profits from last year                 15,286
                                                          36,019
Less Appropriations
    Proposed dividend                          10,000
    General reserve                             1,000
    Foreign exchange                              800     11,800
Unappropriated profits carried to next year               24,219
```

**Figure 1**:   The Conventional Layout

**Stage 1:  Specification and Design of Outputs**

This activity is carried out from the point of view of the *model interpreter(s)*. The model interpreters are the end-users who interpret or use the output of the spreadsheet model for a particular purpose or to make business decisions. The methodology insists on the presentation of outputs on separate worksheets. The output(s) specified would consist of headings, labels and references to the *workings* and *input* sections. These sections would be constructed on separate worksheets later.

Based on the example of a *Trading and Profit and Loss Model*, the model developer would first examine the desired output(s). A typical output structure is shown in *Figure 2*.

**Unappropriated profits carried to next year**

|   | B | C |
|---|---|---|
| 3 | Net profit | ? |
| 4 | *Add* Unappropriated profits from last year | ? |
| 5 | *Less* Appropriations | ? |
| 6 | Unappropriated profits carried to next year | ? |

**Net profit**

|    | B | C |
|----|---|---|
| 11 | Gross Profit | ? |
| 12 | *Less* Expenses | ? |
| 13 | Net profit | ? |

**Figure 2:** Output Structure

If this layout was presented to a group of model developers, who are each asked to independently produce the spreadsheet model, they would come up with different layouts and structures based on experience and personal likes (and dislikes). Adopting the proposed methodology, a group of model developers assigned to independently build the spreadsheet model, should produce a set of structurally identical models.

**Stage 2: Conceptual Design of the Workings Section**

The purpose of constructing the *workings* or *calculations* section is to systematically and methodically perform the interim and final calculations based on and required by the model output(s). In developing the conceptual model, the first step is to identify the highest-level functions or model elements. These take the form of formulae with no dependents. They are therefore not referenced by any other elements within the spreadsheet model. Such functions can be referred to as *root elements* of the model.

The workings section of the spreadsheet model is represented in the form of Jackson structures [Jackson-1975]. The root elements would be placed at the top of the hierarchy, hanging from a box containing the title of the spreadsheet model. The immediate precedents of the root element would then assume their positions just below, adjacent to each other. In the same manner, each element would be positioned just below the model element of which it is a direct precedent. In many spreadsheet models, a root element would represent multiple instances, where each instance corresponds to a different time period, group, category, etc. This is shown as an iteration (appropriately labelled) with the root element appearing below it.

When a top-down approach is adopted without allowing duplication of elements, the initial model could take the form of a graph structure as opposed to the desired tree structure. The purpose of this is to distinctly show instances of multiple dependants of a particular element of the model. This potentially results in a structure as shown in *Figure 3*.

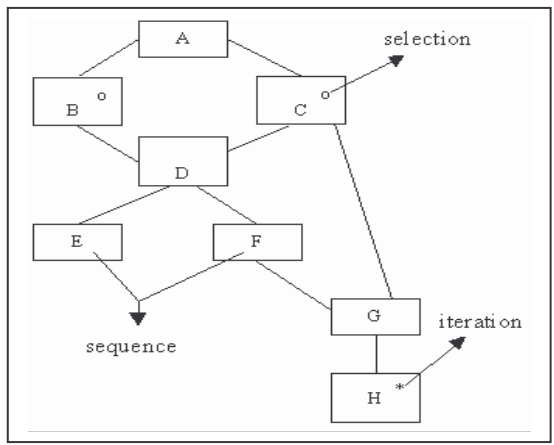

**Figure 3:** The Conceptual Design in Graph-form

Based on the desired outputs for the *Trading and Profit and Loss Account* example shown in *Figure 2*, the only root element that can be identified is **Unappropriated profits carried to next year**, as it is not referenced by any other model element. *Figure 4* presents the conceptual design of the workings section.

**Figure 4:** Conceptual Design of the *Workings* Section

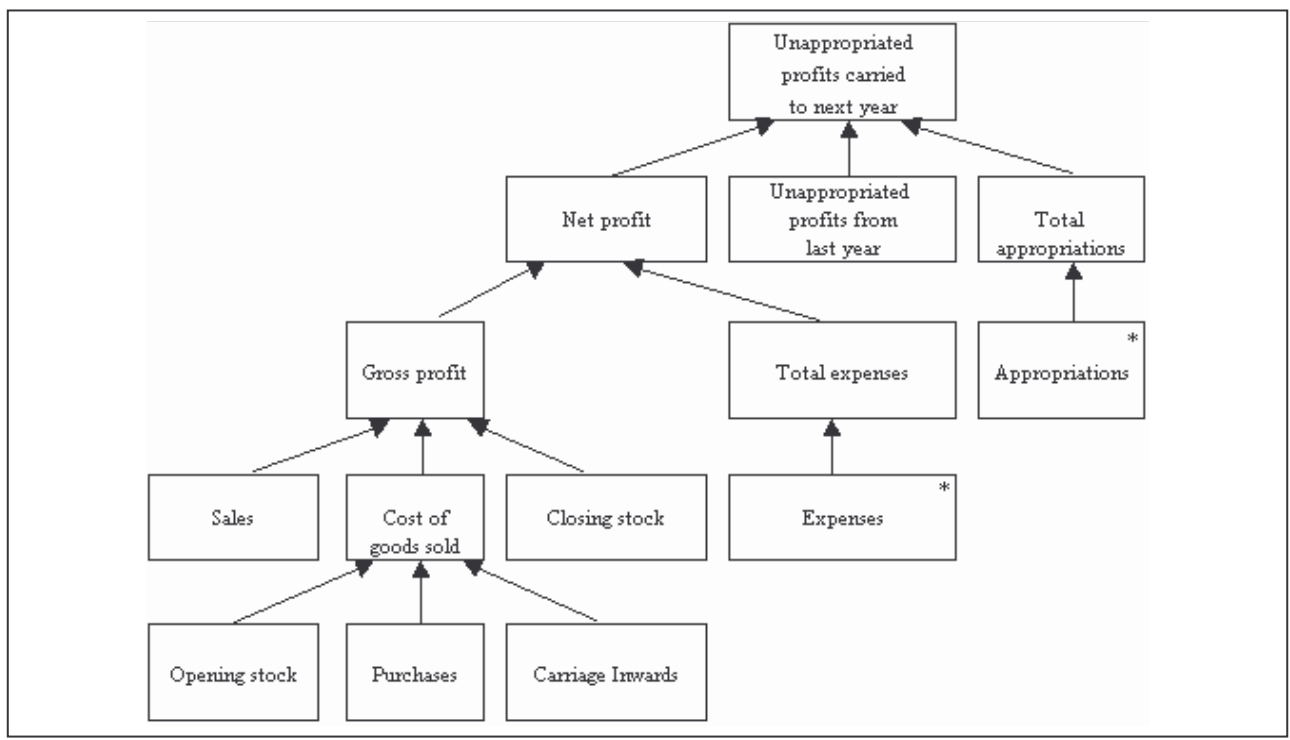

This model distinctly shows the precedents of the various functions. The leaves *expenses* and *appropriations* are represented as iterations in *Figure 4*. This is because each of them refers to a group of related inputs, defined as a range. The elements of a range are always operated on or manipulated as a set rather than individually. This structure should be transformed or resolved into a pure tree structure (if it is not already so) in the next stage.

**Stage 3: Logical Design of the Workings Section**

We have looked at a conceptual design which took the form of a tree. Not all spreadsheet models are of this simple form, but have structure charts in the form of a more general graph. *Figure 3* shows an example of such a chart. The chart is different to that in *Figure 4* in that there is a loop in the relationships connecting **A**, **B, C** and **D**, so that we do not any longer have a tree form. In this chart, data block D contributes to block B and to block C.

We can of course turn the graph into a tree. In order to accomplish this, two important rules have to be observed.

**Rule 1:**

*The initial graph structure is resolved into a tree-structure by duplicating elements with more than one dependant. The precedents of these elements are not included in the model at this stage. This is illustrated in Figure 5.*

Based on *Figure 5*, **D** and **G** are duplicated in order to resolve the graph into a tree structure. The precedents of **D** are not included in the model.

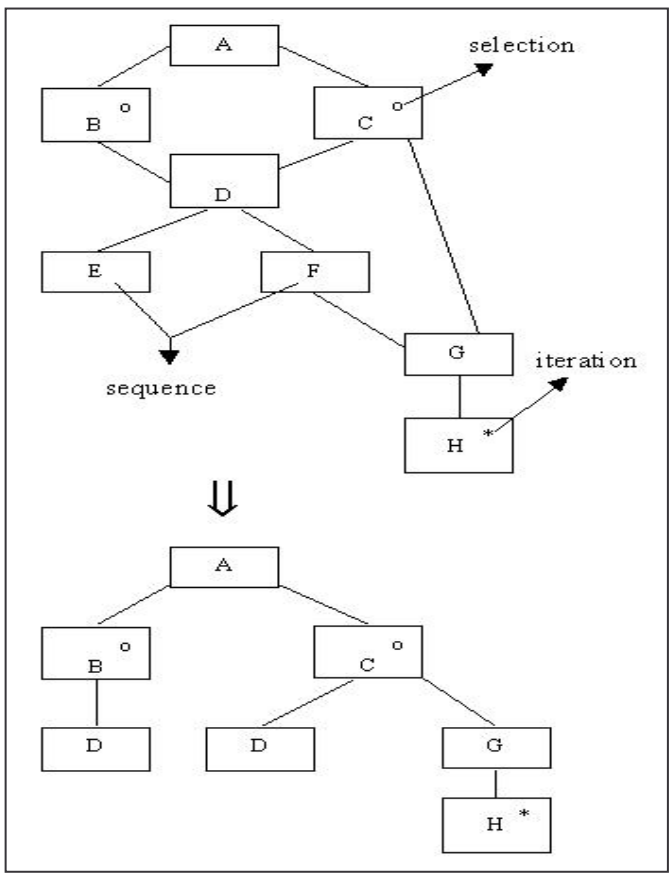

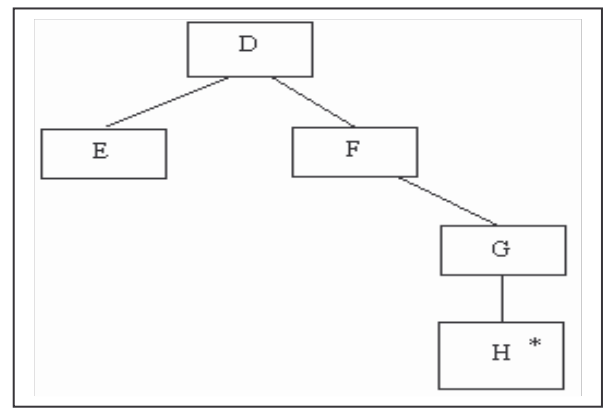

**Figure 5:**  The Logical Design Based on *Rule 1*     **Figure 6:**  The Logical Design Based on *Rule 2*

**Rule 2:**

*If a duplicate element has one or more precedents, it forms a separate module represented as a tree. The module consists of the duplicate element along with its precedents. This process is similar to First Normal Form (1NF) of normalisation in relational database design. This is shown in Figure 6.*

Referring to our Trading and Profit and Loss Account example, the model does not contain any graph sub-structures. Therefore, this stage is not applicable and can be omitted/skipped. In other words, the conceptual design of the workings section also represents its logical design.

**Stage 4: Construction of the Workings Section Structure**

To maintain the structure modelled in the logical design in the spreadsheet view, the indentation principle is used, both on the row labels and on the corresponding values themselves. In fact, we can also insist that the values are indented by assigning a spreadsheet column to each level of indentation. These can be referred to as *virtual columns*. The logical design of the model (represented as Jackson tree-like structures) is systematically mapped onto the physical spreadsheet based on rigorous rules prescribed by the methodology.

The following are the types of elements that can be found in the Jackson structures:

- Iteration (not associated with a single data value)
- Selection (representing mutually exclusive options)
- Function (takes the form of a formula)
- Leaf (reference to an input or input range)
- Constant (reference to an input and denoted using a C)
- Labels (each label is associated with a function, constant, leaf or iteration)
- Module (a branch or sub-structure referenced more than once within the model)

- Reference to a different iteration of the same module (indicated using *indices*)

All *iteration labels* are placed in the same column but are suitably indented to reflect their levels. All *function* and *input labels* are also placed in the same column, adjacent to the column containing iteration labels. They too are indented according to the levels at which they occur in the Jackson structures. The functions corresponding to the labels are built in a set of (virtual) columns adjacent to the column containing the function and input labels. The functions are located in different *virtual columns*, according to their position in the data structures. The term 'virtual columns' is used as the multiple physical columns are viewed as a single *logical* column. As such, each row can only contain exactly one function.

The positioning of functions in the various virtual columns is also consistent with the indentation of their corresponding labels. When these functions appear in different 'virtual columns', the comprehensibility of the model is improved significantly. The precedents of each function become easily identifiable.

**Figure 7:** Organisation of Functions in *Virtual Columns*

Based on *Figure 7*, **A** is a function of **B** and **C**, while **C** is a function of **D** and **G**. The precedents of function **D** are **E** and **F**, whereas the precedents of function **G** are $H_1$, $H_2$ and $H_3$. Referring to the *Trading and Profit and Loss Account* example, the logical design of the workings section of the spreadsheet model is now mapped onto the physical spreadsheet. This is shown in *Figure 8*.

**Figure 8:** *Workings* Section Structure

An asterisk (*) is placed next to a function label (in column A) to denote that the function operates on an input range (defined in the *input* section). The definition of a range in this context is described in the next stage.

**Stage 5: Construction of the Input Section**

There are reasons why cells for data input should be grouped together in an input section, separate from the structured modules described above. One reason is to do with the utmost importance of obtaining accurate data entry. A second reason is that input cells are often referred to by more than one calculated cell. Apart from these reasons, it is also a precaution against any accidental overwriting of formulae. This strategy is similar to the method introduced by DiAntonio [DiAntonio, 1986]. DiAntonio's method advocates the isolation of facts by splitting the spreadsheet into two parts, one for the *facts* and one for the *solution*.

The design of this part of the user interface should be as free from constraints as possible; so as not to hinder the main objective: ease of use and absence of data errors. We are therefore, quite at liberty to put all data input cells into unstructured modules, since there are never any dependencies between them. Any dependency relationship in spreadsheet involves a calculated cell, and either other calculated cells or data input cells. However, they do not exist between data input cells and other data input cells.

Based on the *leaves* identified in the Jackson structures, the *input* section can be created. A problem that can be anticipated at this stage is the difficulty in adding or deleting data from the *input* section while having the changes reflected in the *workings* section. In view of this problem, the methodology requires that a group of related inputs be defined as a range and only the range is referred to in the workings section. A reference to a group of related inputs or an input set (range) is shown in the Jackson structure by a *leaf* represented as an iterated component.

The input section for the *Trading and Profit and Loss Model* can now be created in order to provide the workings section with the values required. This is done on a separate worksheet. The worksheet should be labelled **input**. Based on the logical design for the spreadsheet model, shown in *Figure 4*, the end-leaves can be implemented in an *input* section. This is shown in *Figure 9*.

The input data corresponding to the input groups **expenses (C11 to C18)** and **appropriations (C20 to C22)** are defined as ranges, and assigned the range names *expenses* and *appropriations* respectively.

|    | A                                   | B          | C         |
|----|-------------------------------------|------------|-----------|
| 05 | Sales                               | 135,486.00 |           |
| 06 | Opening stock                       | 40,360.00  |           |
| 07 | Closing stock                       | 52,360.00  |           |
| 08 | Purchases                           | 72,360.00  |           |
| 09 | Carriage inwards                    | 1,570.00   |           |
| 10 | **Expenses**                        |            |           |
| 11 | Salaries                            |            | 18,310.00 |
| 12 | Rates and occupancy                 |            | 4,515.00  |
| 13 | Carriage outwards                   |            | 1,390.00  |
| 14 | Office expenses                     |            | 3,212.00  |
| 15 | Sundry expenses                     |            | 1,896.00  |
| 16 | Depreciation: Buildings             |            | 5,000.00  |
| 17 | Depreciation: Equipment             |            | 9,000.00  |
| 18 | Directors' remuneration             |            | 9,500.00  |
| 19 | **Appropriations**                  |            |           |
| 20 | Proposed dividend                   |            | 10,000.00 |
| 21 | General reserve                     |            | 1,000.00  |
| 22 | Foreign exchange                    |            | 800.00    |
| 23 | Unappropriated profits from last year | 15,286.00 |           |

**Figure 9:** Input Section

**Stage 6: Implementation of Functions and Relationships**

The structured spreadsheet modules represent the calculation or workings section. The structured spreadsheet modules also facilitate auditing and comprehension of the composition/meaning of calculations (expressed as formulae). The various formulae can now be physically implemented or programmed. This stage involves constructing the various formulae and functions required in the workings section. The workings section structure has already been produced and will be used as a basis for the creation of the appropriate functions.

References to inputs are first entered into the relevant cells in the workings section. This includes functions on input ranges, such as **total expenses** and **total appropriations**.

A bottom-up approach is taken in the implementation of formulae and functions in the workings section. *Figures 10 (a)* and *10 (b)* show the final state of the workings section of the *Trading and Profit and Loss* model. In *Figure 10 (a)*, the structure of the underlying functions are shown as entered by the model developer. *Figure 10 (b)*, on the other hand, shows the surface values of the functions based on the current state of inputs.

| | A | B | C | D | E | F |
|---|---|---|---|---|---|---|
| | | £ | £ | £ | £ | £ |
| 06 | Unappropriated profits carried to next year | =(C7+C16)-C17 | | | | |
| 07 | Net profit | | =D8-D15 | | | |
| 08 | Gross Profit | | | =E9-E10+E14 | | |
| 09 | Sales | | | | =Input!B5 | |
| 10 | Cost of goods sold | | | | =SUM(F11:F13) | |
| 11 | Opening stock | | | | | =Input!B6 |
| 12 | Add Purchases | | | | | =Input!B8 |
| 13 | Add Carriage inwards | | | | | =Input!B9 |
| 14 | Closing stock | | | | =Input!B7 | |
| 15 | Total expenses * | | | =SUM(Expenses) | | |
| 16 | Add Unappropriated profits from last year | | =Input!B23 | | | |
| 17 | Less Total appropriations * | | =SUM(Appropriations) | | | |

**Figure 10 (a):** *Workings* Section

| | A | B | C | D | E | F |
|---|---|---|---|---|---|---|
| | | £ | £ | £ | £ | £ |
| 06 | Unappropriated profits carried to next year | 24,219 | | | | |
| 07 | Net profit | | | 20,733 | | |
| 08 | Gross Profit | | | | 73,556 | |
| 09 | Sales | | | | | 135,486 |
| 10 | Cost of goods sold | | | | | 114,290 |
| 11 | Opening stock | | | | | 40,360 |
| 12 | Add Purchases | | | | | 72,360 |
| 13 | Add Carriage inwards | | | | | 1,570 |
| 14 | Closing stock | | | | | 52,360 |
| 15 | Total expenses * | | | | 52,823 | |
| 16 | Add Unappropriated profits from last year | | | 15,286 | | |
| 17 | Less Total appropriations * | | | 11,800 | | |

**Figure 10 (b):** *Workings* Section

Based on *Figures 10 (a)* and *10 (b)* it can be noticed that both the semantics and the data are clarified in this layout. For example, we can see straight away on the semantic level that *Unappropriated profits carried to next year* is derived from three figures: *Net Profit*, *Unappropriated profits from last year* and *Total appropriations*.

On the data level we see that **24,219** is made up from **20,733**, **15,286** and **11,800**. Likewise, we see immediately (from the asterisk **\***) that *Total expenses* references an input range from the input section. Notice also that columns in the spreadsheet show figures on the same semantic level, enabling valid comparisons between figures to be made. For example, column C shows *net profit*, *unappropriated profits from last year* and *total appropriations*. These figures give a valid impression of the state of the *Trading and Profit and Loss Account* at this level of detail. If we were to include a figure from a different level, e.g. *purchases* (from column F), it would confuse the picture, since it has already been included in net profit.

Referring to *Figures 10 (a)* and *10 (b)*, it is beyond any doubt that the use of *indentation* and *virtual columns* make it far more straight-forward to make sense of and comprehend the composition of functions. However, the fact that references to data and other formulae within a particular formula take the form of cell addresses rather than meaningful labels is not entirely desirable.

In order to enhance the comprehensibility of formulae, cell addresses should be replaced with meaningful labels so that formulae are expressed in natural language form. Based on the *Trading and Profit and Loss Account* example, meaningful names would first be assigned to every piece of input data. The exception to this rule applies to a data value which is part of a related set of data that is always treated and operated on as a set, in which case it will be defined as a range along with the other related inputs. If every piece of input data in the input section is given a unique name, the workings section would now appear as shown in *Figure 10 (c)*.

**Figure 10 (c):** *Workings* Section

|    | A | B | C | D | E | F |
|----|---|---|---|---|---|---|
| 05 |   | £ | £ | £ | £ | £ |
| 06 | Unappropriated profits carried to next year | =(C7+C16)-C17 | | | | |
| 07 | Net profit | | =D8-D15 | | | |
| 08 | Gross Profit | | | =E9-E10+E14 | | |
| 09 | Sales | | | | =SalesIn | |
| 10 | Cost of goods sold | | | | =SUM(F11:F13) | |
| 11 | Opening stock | | | | | =OpeningStockIn |
| 12 | *Add* Purchases | | | | | =PurchasesIn |
| 13 | *Add* Carriage inwards | | | | | =CarriageInwardsIn |
| 14 | Closing stock | | | | =ClosingStockIn | |
| 15 | Total expenses * | | | =SUM(ExpensesIn) | | |
| 16 | *Add* Unappropriated profits from last year | | =UnappropriatedProfitsFromLastYearIn | | | |
| 17 | *Less* Total appropriations * | | =SUM(AppropriationsIn) | | | |

This technique should be applied to all elements of the workings section. Every function/formula should be assigned a name so that meaningful names instead of cell addresses can be used for references within formulae in the *workings* and *output* sections. This is shown in *Figures 10 (d)* and *11 (c)*.

|    | A | B | C | D | E | F |
|----|---|---|---|---|---|---|
| 05 |   | £ | £ | £ | £ | £ |
| 06 | Unappropriated profits carried to next year | =(NetProfit+UnappropriatedProfitsFromLastYear)-TotalAppropriations | | | | |
| 07 | Net profit | | =GrossProfit-TotalExpenses | | | |
| 08 | Gross Profit | | | =Sales-CostOfGoodsSold+ClosingStock | | |
| 09 | Sales | | | | =SalesIn | |
| 10 | Cost of goods sold | | | | =SUM(OpeningStock:CarriageInwards) | |
| 11 | Opening stock | | | | | =OpeningStockIn |
| 12 | *Add* Purchases | | | | | =PurchasesIn |
| 13 | *Add* Carriage inwards | | | | | =CarriageInwardsIn |
| 14 | Closing stock | | | | =ClosingStockIn | |
| 15 | Total expenses * | | | =SUM(ExpensesIn) | | |
| 16 | *Add* Unappropriated profits from last year | | =UnappropriatedProfitsFromLastYearIn | | | |
| 17 | *Less* Total appropriations * | | =SUM(AppropriationsIn) | | | |

**Figure 10 (d):** *Workings* Section

**Stage 7: Completion of the Output Section**

This stage brings the spreadsheet model development process to a conclusion. References to corresponding functions in the workings section can at this stage be entered into the relevant cells of the output section. The final state of the output section is shown in *Figures 11 (a), 11 (b)* and *11 (c)*.

|    | B                                              | C      |
|----|------------------------------------------------|--------|
| 01 | **Unappropriated profits carried to next year** |        |
| 02 |                                                |        |
| 03 | Net profit                                     | 20,733 |
| 04 | *Add* Unappropriated profits from last year    | 15,286 |
| 05 | *Less* Appropriations                          | 11,800 |
| 06 | **Unappropriated profits carried to next year** | 24,219 |
| 07 |                                                |        |
| 08 |                                                |        |
| 09 | **Net profit**                                 |        |
| 10 |                                                |        |
| 11 | Gross Profit                                   | 73,556 |
| 12 | *Less* Expenses                                | 52,823 |
| 13 | **Net profit**                                 | 20,733 |
| 14 |                                                |        |

**Figure 11 (a):** *Output* Section

|    | B                                              | C               |
|----|------------------------------------------------|-----------------|
| 01 | **Unappropriated profits carried to next year** |                 |
| 02 |                                                |                 |
| 03 | Net profit                                     | =Workings!C7    |
| 04 | *Add* Unappropriated profits from last year    | =Workings!C16   |
| 05 | *Less* Appropriations                          | =Workings!C17   |
| 06 | **Unappropriated profits carried to next year** | =Workings!B6    |
| 07 |                                                |                 |
| 08 |                                                |                 |
| 09 | **Net profit**                                 |                 |
| 10 |                                                |                 |
| 11 | Gross Profit                                   | =Workings!D8    |
| 12 | *Less* Expenses                                | =Workings!D15   |
| 13 | **Net profit**                                 | =Workings!C7    |
| 14 |                                                |                 |

**Figure 11 (b):** *Output* Section

|    | B                                              | C                                      |
|----|------------------------------------------------|----------------------------------------|
| 01 | **Unappropriated profits carried to next year** |                                        |
| 02 |                                                |                                        |
| 03 | Net profit                                     | =NetProfit                             |
| 04 | *Add* Unappropriated profits from last year    | =UnappropriatedProfitsFromLastYear     |
| 05 | *Less* Appropriations                          | =TotalAppropriations                   |
| 06 | **Unappropriated profits carried to next year** | =UnappropriatedProfitsCarriedToNextYear |
| 07 |                                                |                                        |
| 08 |                                                |                                        |
| 09 | **Net profit**                                 |                                        |
| 10 |                                                |                                        |
| 11 | Gross Profit                                   | =GrossProfit                           |
| 12 | *Less* Expenses                                | =TotalExpenses                         |
| 13 | **Net profit**                                 | =NetProfit                             |

**Figure 14 (c):** *Output* Section

# 3. EVALUATION OF THE STRUCTURED METHODOLOGY

## 3.1 Introduction

In order to evaluate the effectiveness of the proposed methodology, a series of well-organised experiments were undertaken. An analysis of the results of these experiments would revealed the methodology's potential for integrity control of spreadsheet models. A major problem encountered was to persuade certain groups of spreadsheet users, especially those in industry (as opposed to academia) to take part in the trials. Therefore, the selection of user groups involved consideration of various factors such as circumstances, experience and various constraints. Two different strategies are formulated to evaluate the quality of the proposed methodology for spreadsheet model development.

**User Groups or Participants**

Ideally, the methodology should be tested on spreadsheet users, of varying levels of spreadsheet literacy, in both business and academia. Past experiments on spreadsheet errors have involved different types of users, from experienced spreadsheet developers from industry to novice spreadsheet students. It has been impossible for the authors to obtain consent to conduct trials with users in business organisations due to various reasons e.g. the time commitment, the problem of confidentiality of client data, the difficulty of obtaining a cohort of users all working with the same model at the same time, as well as the difficulty of obtaining a sufficiently large cohort to produce statistically significant results.

Referring to past experiments undertaken, it is found that most of the participants of such tests were students at an institution associated with the author(s). In most cases where the subjects were industry or commercial users, the experiment was either conducted by the particular organisation or the information derived from the normal operations of the organisation, published by the company itself.

Three different groups of students at a University were selected as participants for the experiments. They were as follows:

- Undergraduates
- Post-graduate students
- Students on a short course designed primarily for professionals in industry.

**Types of Errors**

Ideally, the tests should demonstrate the capacity of the proposed structured methodology to address all types of spreadsheet errors. The taxonomy or classification of spreadsheet errors [Rajalingham et al, 2000] is used as a basis for organising tests for as many different types of errors as possible.

**Spreadsheet Models**

The spreadsheet models selected and used for experimental purposes should be common business and financial models. The models should address the different features of the proposed methodology. Moreover, the models should have the capacity to be used to test for as many different types of errors as possible.

The spreadsheet models selected for the experiments are as follows:

- A Trading and Profit and Loss Account for a particular year [Wood-96]
- A Trading and Profit and Loss Account for several years [Wood-96]
- A Post-tax Income Distribution Model [Slater-90]
- Another common business model.

### 3.2 The Evaluation Strategies

**Error Prevention**

The first strategy for testing the quality of the proposed methodology is based on error prevention. It involves comparing the occurrence of errors in spreadsheet models developed based on the proposed methodology to the occurrence of errors in models built using conventional unstructured methods. The aim of this strategy is to establish whether or not there is a material difference in error rates between spreadsheet models produced using the two different approaches. The hypothesis is that users commit significantly fewer errors by adopting the proposed structured methodology. The first experiment is based on this strategy while the subsequent three experiments are based on a different strategy (error detection).

**Error Detection**

The second strategy for evaluating the effectiveness of the proposed methodology is based on error detection. It involves comparing the probability of detecting errors in spreadsheet models developed based on the proposed methodology to the probability of detecting errors in models constructed based on conventional unstructured methods. Errors are deliberately seeded into the spreadsheet models. The aim of this strategy is to establish whether or not there is a significant difference in the probabilities of error detection between spreadsheet models produced using the two different approaches. The hypothesis is that users are able to identify significantly more errors seeded into a model developed using the proposed structured methodology. This is a reflection of its comprehensibility. This is particularly important for audit, review and update purposes. Apart from the first experiment, the subsequent four experiments are based on this strategy.

### 3.3 The Experiments and their Results

**Experiment 1**

This experiment was carried out in two different stages, each involving two groups of students at a University. The purpose of the experiment was to compare two different approaches to the development of a single-module spreadsheet model. The first approach was based on conventional unstructured methods for spreadsheet model development while the second approach was based on the proposed structured methodology. This experiment was based on the first testing strategy, described earlier. The spreadsheet model used was based on a *Trading and Profit and Loss Account* for a particular year [Wood-96].

*Stage 1*

The first stage of the experiment involved the development of a spreadsheet model without any guidance or support. Subjects were given the desired output of the model as shown in *Figure 3*. In order to create the spreadsheet model based on the required output, they were provided with all the formulae needed. They had to employ suitable methods based on personal experience or discretion, and carry out the exercise independently. A total of **42 post-graduate students** and **26 short course students** (most of whom were professionals in industry) took part in this experiment.

The first test was carried out on a group of **22 post-graduate students**. The students were pursuing a taught masters programmme. Most of them had graduated in other disciplines and had limited prior knowledge of information systems. Each participant had to build the same spreadsheet model on two different occasions. The purpose of having the participants re-build the same model was so that it can be used as a control in the experiment.

The second test was performed on a group of **12 short course students**. Most of the students were employed on a full-time basis in industry. Each participant had to build the spreadsheet model without having had a lesson on the proposed methodology.

*Stage 2*

The second stage of the experiment involved the development of the same spreadsheet model based on a *Trading and Profit and Loss Account*. However, before they carried out the exercise, the students were given a lesson on employing the proposed methodology for structuring and building a single-module spreadsheet model.

The first test was carried out on a group of **20 post-graduate students**. The students were also pursuing a taught masters programmme. Each participant had to first build the spreadsheet model using a method they were familiar with. This was not based on any structured methodologies. The purpose of this exercise was to make sure that the errors committed by this group of students were in fact consistent with those produced by the previous group. The group was subsequently given a lesson on using the proposed methodology to construct a single-module spreadsheet model. They were then asked to re-construct the spreadsheet model based on the proposed methodology.

The second test was conducted on a group of **14 short course students**. This was a different group of students but were pursuing the same short course. Moreover, they had a similar background, in that they were also mainly professionals in industry. The participants were asked to create the spreadsheet model, having had a lesson on building spreadsheet models using the proposed methodology.

*Results*

|  | Test 1a | Test 1b | Test 2 |
| --- | --- | --- | --- |
| **Stage 1** | | | |
| Subject | Post-graduate Students | | Short Course Students |
| Sample Size | 22 | 22 | 12 |
| Mean Number of Errors | **3.9** | **3.8** | **3.3** |
| **Stage 2** | | | |
| Subject | Post-graduate Students | | Short Course Students |
| Sample Size | 20 | 20 | 14 |
| Mean Number of Errors | **3.6** | **1.4** | **1.2** |

The errors include both quantitative and qualitative errors [Rajalingham et al, 2000]

**Experiment 2**

This experiment was based on the second evaluation strategy (error detection) and carried out in two stages. A total of **104 undergraduates** took part in this experiment. The students were in two different groups. Both groups had to detect a total of 12 errors that had been seeded into a spreadsheet model. They were given the same amount of time to complete the exercise. The model was based on a *Trading and Profit and Loss Account* for several years. However, there was a fundamental difference between the layout or structure of the model used by the first group and the model used by the second group.

*Stage 1*

The first group consisted of **55 students** and were presented a spreadsheet model in a conventional layout. Their task was to identify the twelve errors that had been seeded into the model. They were not aware of how many errors there were in the model.

*Stage 2*

The second group, on the other hand, was made up of **49 students**. This group was working on the same model but it was structured based on the proposed methodology. The same errors had been seeded into this model as well, and group members had to independently detect them. gain they were unaware of the number of errors. They were given a brief and general lesson on how to interpret a spreadsheet model based on the proposed methodology without any references to the particular model used.

*Results*

| Stage 1 | |
|---|---|
| Subject | Undergraduates |
| Sample Size | 55 |
| **Mean Errors Detected (%)** | **32.5** |
| Stage 2 | |
| Subject | Undergraduates |
| Sample Size | 49 |
| **Mean Errors Detected (%)** | **71.7** |

**Experiment 3**

This experiment was based on the second evaluation strategy (error detection) and carried out in two stages. A total of **41 post-graduate students** and **23 short course students** participated in this experiment. Two identical tests were performed in each stage. Each test involved a different subset of students. Therefore 4 groups of subjects had to detect a total of 10 errors that had been seeded into a spreadsheet model. The model used in the first stage had a different structure/layout to the model used in the second stage. All participants were given the same amount of time to complete the exercise. The model used in this experiment was based on a *Post-tax Income Distribution Model* [Slater-90]. The original model was modified slightly to decrease its size.

*Stage 1*

In the first stage of the experiment, the spreadsheet model was presented based on the original (conventional) layout. They had to identify a total of 10 errors that had been seeded into the model. The first test involved a group of **19 post-graduate students** while the second test was conducted on a group of **11 short course students**.

*Stage 2*

In the second stage of the experiment, the spreadsheet model was re-designed and re-structured according to the proposed methodology. The same 10 errors were then deliberately seeded into the model. The participants of the experiment at this stage were given a brief and general lesson on how to interpret a spreadsheet model based on the proposed methodology without any references to the particular model used.

The first test was performed on a group of **22 post-graduate students** while the second test involved a group of **12 short course students**.

| | Test 1 | Test 2 |
|---|---|---|
| Stage 1 | | |
| Subject | Post-graduate Students | Short Course Students |
| Sample Size | 19 | 11 |
| **Mean Errors Detected (%)** | **22.8** | **24.0** |
| Stage 2 | | |
| Subject | Post-graduate Students | Short Course Students |
| Sample Size | 22 | 12 |
| **Mean Errors Detected (%)** | **52.4** | **58.1** |

*Results*

**Experiment 4**

This experiment was very similar to the previous experiment. The only difference was that a different spreadsheet model was used. However, this was also a common business model. The model was simplified and its size reduced to make it less time-consuming to work on. The experiment was carried out in two stages and involved a total of **44 post-graduate students** and **23 short course students**. Two identical tests were performed in each stage. Each test involved a different subset of students. The task of the 4 groups of participants was to detect a total of 10 errors that had been seeded into the spreadsheet model. All participants were given the same amount of time to complete the exercise.

*Stage 1*

In the first stage of the experiment, the spreadsheet model was presented based on the original (conventional) layout. The first test involved a group of **24 post-graduate students** while the second test was conducted on a group of **12 short course students**.

*Stage 2*

In the second stage of the experiment, the spreadsheet model was re-designed and re-structured according to the proposed methodology. They same 10 errors were then seeded into the model. As done in the previous experiment, the students taking part in the experiment at this stage were given a brief and general lesson on how to interpret a spreadsheet model based on the proposed methodology without any references to the particular model used. The first test was performed on a group of **20 post-graduate students** while the second test involved a group of **11 short course students**.

*Results*

|  | Test 1 | Test 2 |
|---|---|---|
| **Stage 1** | | |
| Subject | Post-graduate Students | Short Course Students |
| Sample Size | 24 | 12 |
| Mean Errors Detected (%) | **44.3** | **36.5** |
| **Stage 2** | | |
| Subject | Post-graduate Students | Short Course Students |
| Sample Size | 20 | 11 |
| Mean Errors Detected (%) | **67.8** | **78.1** |

**4. CONCLUSION**

The suitability of a methodology based on Jackson-like structures for spreadsheet modelling has been investigated. It appears that there are several possible advantages to the adoption of a structured method based on a Jackson data-oriented approach. These advantages may be summarised as follows [Knight et al, 2000]:

- a clear modularisation principle
- a top-level overview of module structure
- a structured indented format to the layout of module
- the possibility of automatic structuring of existing spreadsheets

The proposed methodology imposes a strict discipline in the process of spreadsheet development using software engineering principles. This reduces the occurrence of errors as spreadsheet models are designed and constructed in a structured and organised manner. The methodology distinctly describes a technique for modelling the spreadsheet problem and subsequently mapping the design onto the physical spreadsheet according to prescribed rules and a structured algorithm. The spreadsheet model is organised in a form which facilitates understanding and interpretation of the model in an unambiguous way. It is also appropriately decomposed into modules. This reduces the occurrence of most types of errors and increases the probability of detecting errors which are already present in the spreadsheet models.

In order to assess and establish the quality of the methodology, five different experiments have been carried out. The results of these experiments have been analysed and they are presented in this paper. The results of the series of four experiments conducted provide substantial evidence of the methodology's potential for controlling the integrity and improving the comprehensibility of spreadsheet models. A more detailed version of the complete set of experiments and analysis of their results will be published in due course.